# Acoustically mediated long-range interaction among multiple spherical particles exposed to a plane standing wave

Shenwei Zhang,[1] Chunyin Qiu,[1*] Mudi Wang,[1] Manzhu Ke,[1] and Zhengyou Liu[1,2,3]

[1]Key Laboratory of Artificial Micro- and Nano-structures of Ministry of Education and School of Physics and Technology, Wuhan University, Wuhan 430072, China

[2]Institute for Advanced Studies, Wuhan University, Wuhan 430072, China

[3]Department of Physics, South China University of Technology, Guangzhou 510641, China

In this work, we study the acoustically mediated interaction forces among multiple well-separated spherical particles trapped in the same node or antinode plane of a standing wave. An analytical expression of the acoustic interaction force is derived, which is accurate even for the particles beyond the Rayleigh limit. Interestingly, the multi-particle system can be decomposed into a series of independent two-particle systems described by pairwise interactions. Each pairwise interaction is a long-range interaction, as characterized by a soft oscillatory attenuation (at the power exponent of $n = -1$ or $-2$). The vector additivity of the acoustic interaction force, which is not well expected considering the nonlinear nature of the acoustic radiation force, is greatly useful for exploring a system consisting of a large number of particles. The capability of self-organizing a big particle cluster can be anticipated through such acoustically controllable long-range interaction.

*Author to whom correspondence should be addressed. Email: cyqiu@whu.edu.cn

**I. Introduction**

After the pioneering discovery of optical trapping [1,2], Burns et al. [3] observed a series of optically-induced bound states between two polystyrene spheres. Different from the optical trapping generated by the gradient of an external field directly, the optical binding effect occurs even in a plane wave field without any gradient. Since then, extensive efforts have been devoted to the light-mediated particle-particle interaction [4-12]. The optical binding force alternates between attraction and repulsion with the inter-particle separation, which leads to a light-mediated self-organization of particle clusters. The interaction between objects can also be induced by acoustic waves, as reported theoretically [13-24] and experimentally [25-28]. Comparing with its optical counterpart, less progress has been made on the sound-mediated interaction. To the best of our knowledge, most of the current theoretical studies focus on the acoustic interaction between two objects [14-20].

The acoustic radiation force (ARF), as a second order quantity of the acoustic field, relies on an accurate calculation of the self-consistent field distribution in the presence of the particles. The self-consistent sound field can be solved by various techniques, such as the finite-element method, the finite difference time domain simulation, and the multiple-scattering theory (MST). For spherical particles, the MST approach has been proved to be the most efficient since the sound field can be precisely captured by a finite number of spherical basis functions. It has been successfully developed to calculate the acoustic interaction between two spherical particles [14,15]. Although a system involving multiple particles [21-23] can also be handled, the MST method becomes inefficient as the growth of the particle number, owing to the overloaded memory and time consumption. Recently, for the particles much smaller than the acoustic wavelength (i.e., Rayleigh particles), a simple formula of the pairwise interaction force has been derived based on a scalar potential theory [24]. Together with a mean-field approximation, this approach is further demonstrated to be powerful in treating a great number of the Rayleigh particles. It is worth pointing out that, the acoustic interaction among the Rayleigh particles is very weak, which could be hidden in the gradient force induced by a tiny defect of the external sound field.

Based on a single prior scattering approximation, here we present a theoretical study on the sound-mediated interactions among the multiple spherical particles distributed sparsely in the same node or antinode plane of a plane standing wave (PSW) field. Particularly, we focus on the particle size comparable with the acoustic wavelength, in which the acoustic interaction is anticipated to be much stronger than that in the Rayleigh situation. We derive a concise form for the total interaction force exerting on any given particle. It consists of a series of independent pairwise interactions between the particle and the others. Therefore, the multi-particle system is reduced into a two-body problem, which greatly simplifies the computation comparing with the rigorous MST method. Interestingly, we find that each pairwise interaction is a conservative force and thus the whole system could be described by a potential energy. This result is unusual for the particle beyond the Rayleigh limit. Besides, the pairwise force is oscillatory at a factor of cosine function and decays at a power exponent of $n = -1$ or $-2$. The oscillatory long-range interaction, controlled by the external sound field, could facilitate the self-organization of a two-dimensional (2D) cluster involving a large number of particles.

The remainder of this paper is organized as follows. In Sec. II, an analytical formula responsible for the long-distance acoustic interaction is derived, followed by a detailed discussion on the formula. In Sec. III we check first the accuracy of our formula by some few-particle systems, and then study the system involving a large number of particles. Finally, a brief summary is made in Sec. IV.

## II. Theoretical derivation of the long-range acoustic interaction

PSW is a good external field for exploring the acoustically mediated particle-particle interaction. The node or antinode plane forms a natural potential well to confine the particles tightly in the same plane without any transversal field gradient. Assume that the acoustic PSW field is described by $\Psi_{\text{in}}^{(e)}(\mathbf{r}) = \Psi_0(e^{ikz} + e^{-ikz})$, where $\Psi_0$ characterizes the amplitude of the PSW field. In this case, the interaction force survives only in the $xy$ plane owing to the symmetry of the external field.

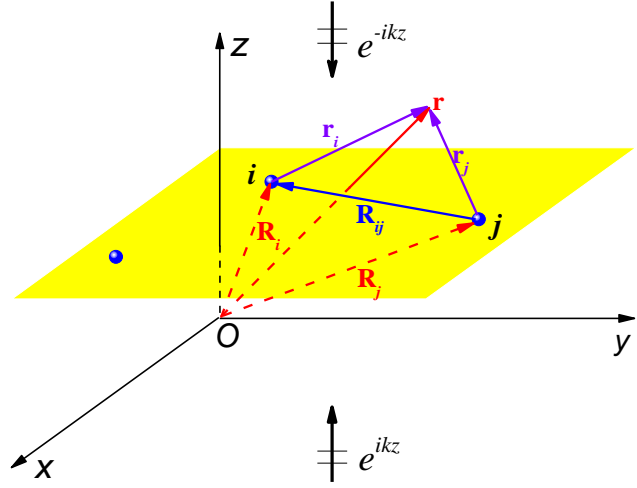

FIG. 1. (Color online) Spherical particles distributed in the same node or antinode plane of an acoustic PSW field consisting of two counter-propagating plane waves. Each color arrow labels a position vector measured from the origin or a given sphere center.

As schematically depicted in Fig. 1, for an arbitrary spherical particle $i$, the total velocity potential function can be written as a superposition of the incident wave $\Psi_{\text{in}}(\mathbf{r})$ and the scattering wave $\Psi_{\text{sc}}(\mathbf{r})$, i.e.,

$$\Psi_{\text{in}}(\mathbf{r}) = \sum_{lm} a^i_{l,m} j_l(kr_i) Y_{l,m}(\hat{r}_i), \qquad (1a)$$

$$\Psi_{\text{sc}}(\mathbf{r}) = \sum_{lm} b^i_{l,m} h_l(kr_i) Y_{l,m}(\hat{r}_i), \qquad (1b)$$

where $k$ is the wavenumber in the host fluid, the vector $\mathbf{r}_i = r_i \hat{r}_i$ refers to any probe position $\mathbf{r}$ measured from the center of a given sphere $i$, and $\{j_l\}$, $\{h_l\}$ and $\{Y_{l,m}\}$ are the spherical Bessel functions, the spherical Hankel functions of the first kind and the spherical harmonics, respectively.

The incident coefficient $a^i_{lm}$ can be decomposed into two parts,

$$a^i_{l,m} = a^{i(e)}_{l,m} + \sum_{j \neq i} a^{i(j)}_{l,m}. \qquad (2)$$

The first part $a^{i(e)}_{l,m}$ stems from the external incidence (PSW field here),

$$a_{l,m}^{i(e)} = \Psi_0 4\sqrt{(2l+1)\pi}\delta_{m0}\cos\left(\alpha + \frac{l\pi}{2}\right), \tag{3}$$

with $\delta_{m0}$ being a Kronecker delta. The value of $\alpha$ depends on the location of the particle ensemble: $0$ for the antinode plane and $\pi/2$ for the node plane of the acoustic PSW field. The second part stems from the scattering of the other particles $j \neq i$. Each component $a_{l,m}^{i(j)}$ can be derived according to the addition theorem of the spherical functions [29], i.e.,

$$a_{l,m}^{i(j)} = \sum_{l'm'} b_{l',m'}^{j} G_{l'm'lm}^{ij}. \tag{4}$$

Here $G_{l'm'lm}^{ij} = 4\pi \sum_{l''} i^{l+l''-l'} C_{lml''(m'-m)}^{l'm'} h_{l''}(kR_{ij}) Y_{l'',m'-m}(\hat{R}_{ij})$ with the coefficient $C_{lml''(m'-m)}^{l'm'} = \int_0^{2\pi}\int_0^{\pi} Y_{l',m'}(\theta,\phi) Y_{l,m}^*(\theta,\phi) Y_{l'',m'-m}^*(\theta,\phi) \sin\theta d\theta d\phi$. $\mathbf{R}_{ij} = R_{ij}\hat{R}_{ij}$ represents the position vector of the spherical center $i$ measured from the spherical center $j$.

For a homogeneous spherical particle, the incident and scattering coefficients can be related by a diagonal scattering matrix $\{t_l^i \delta_{ll'}\}$, i.e., $b_{l,m}^i = t_l^i a_{l,m}^i$, according to the continuum boundary condition on the surface of the particle. Substituting this relation into Eq. (4) and using Eq. (2), we obtain a linear equation system

$$\sum_{jl'm'} \left(\delta_{ij}\delta_{ll'}\delta_{mm'} - t_{l'}^j G_{l'm'lm}^{ij}\right) a_{l',m'}^j = a_{l,m}^{i(e)}. \tag{5}$$

Solving this linear problem gives the incident coefficient $a_{l,m}^i$ and further gives the total self-consistent acoustic field according to Eq. (1). In this procedure, multiple scatterings among the particles are taken into account rigorously. However, the computational time and memory cost increase rapidly with the growth of the particle number. If the particles are distributed sparsely $kR_{ij} \gg 1$, a single prior scattering approximation can be used to avoid solving the linear problem. Now the scattering contribution from the sphere $j$ to $i$ is dominated by a single prior scattering event, i.e., $b_{l,m}^j \approx t_l^j a_{l,m}^{j(e)}$, which finally gives rise to

$$a_{l,m}^{i(j)} \approx \sum_{l'm'} t_{l'}^{j} a_{l',m'}^{j(e)} G_{l'm'lm}^{ij}. \tag{6}$$

Combining the Eqs. (1), (2) and (6) with the relation $b_{l,m}^{i} = t_{l}^{i} a_{l,m}^{i}$, we can obtain the field distribution in the presence of particles. The ARF exerted on any particle $i$ can be calculated by integrating the time-averaged radiation stress tensor $\langle \overleftrightarrow{S} \rangle$ over an arbitrary surface $S$ enclosing the particle, i.e.,

$$\mathbf{F}^{i} = \oiint_{S} \langle \overleftrightarrow{S} \rangle \cdot d\mathbf{A}. \tag{7}$$

Specifically, the time-averaged radiation stress tensor $\langle \overleftrightarrow{S} \rangle$ can be written as

$$\langle \overleftrightarrow{S} \rangle = \frac{1}{2} \rho_0 \text{Re}(\mathbf{v}^* \mathbf{v}) - \left( \frac{\rho_0 |\mathbf{v}|^2}{4} - \frac{|p|^2}{4\rho_0 c_0^2} \right) \overleftrightarrow{I}, \tag{8}$$

with $\overleftrightarrow{I}$ being a unit tensor, $\rho_0$ and $c_0$ being the static mass density and sound velocity of the fluid background, and $\mathbf{v}$ and $p$ denoting the first-order velocity and pressure fields, respectively. If an infinitely large contour is selected, the integral form of the ARF $\mathbf{F}^{i} = (F_x^{i}, F_y^{i})$ can be rewritten as a series [30,31]

$$F_x^{i} + iF_y^{i} = \frac{i\rho_0}{4} \sum_{lm} [\mu_{l+1,m-1}(2b_{l+1,m-1}^{i} b_{l,m}^{i*} + b_{l+1,m-1}^{i} a_{l,m}^{i*} + a_{l+1,m-1}^{i} b_{l,m}^{i*})$$

$$+ \mu_{l+1,-m-1}(2b_{l,m}^{i} b_{l+1,m+1}^{i*} + b_{l,m}^{i} a_{l+1,m+1}^{i*} + a_{l,m}^{i} b_{l+1,m+1}^{i*})], (9)$$

where the coefficient $\mu_{l,m} = \sqrt{[(l-m)(l-m-1)]/[(2l-1)(2l+1)]}$. Using the relation $b_{l,m}^{i} = t_{l}^{i} a_{l,m}^{i}$, Eq. (9) can be reorganized into

$$F_x^{i} + iF_y^{i} = \frac{i\rho_0}{4} \sum_{lm} [T_{l}^{i} \mu_{l+1,m-1} a_{l,m}^{i*} a_{l+1,m-1}^{i} + T_{l}^{i*} \mu_{l+1,-m-1} a_{l,m}^{i} a_{l+1,m+1}^{i*}], \tag{10}$$

where the factor $T_{l}^{i} = 2t_{l+1}^{i} t_{l}^{i*} + t_{l+1}^{i} + t_{l}^{i*}$ characterizes the scattering property of the particle $i$. Substituting Eq. (2) into Eq. (10), the acoustic interaction can be divided into three parts

$$F_x^{i} + iF_y^{i} = \frac{i\rho_0}{4}(Q_e + Q_s + Q_c), \tag{11}$$

where

$$Q_e = \sum_{lm} \left[ T_l^i \mu_{l+1,m-1} a_{l,m}^{i(e)} a_{l+1,m-1}^{i(e)} + T_l^{i*} \mu_{l+1,-m-1} a_{l,m}^{i(e)} a_{l+1,m+1}^{i(e)} \right], \quad (12a)$$

$$Q_s = \sum_{lm} \sum_{j \neq i, h \neq i} \left[ T_l^i \mu_{l+1,m-1} a_{l,m}^{i(j)*} a_{l+1,m-1}^{i(h)} + T_l^{i*} \mu_{l+1,-m-1} a_{l,m}^{i(j)} a_{l+1,m+1}^{i(h)*} \right], \quad (12b)$$

$$Q_c = \sum_{lm} \sum_{j \neq i} [T_l^i \mu_{l+1,m-1} \left( a_{l,m}^{i(e)} a_{l+1,m-1}^{i(j)} + a_{l+1,m-1}^{i(e)} a_{l,m}^{i(j)*} \right)$$

$$+ T_l^{i*} \mu_{l+1,-m-1} \left( a_{l,m}^{i(e)} a_{l+1,m+1}^{i(j)*} + a_{l+1,m+1}^{i(e)} a_{l,m}^{i(j)} \right)]. \quad (12c)$$

The first term $Q_e$ involves only the external field $a_{l,m}^{i(e)}$, which is always zero because $a_{l,m}^{i(e)} a_{l+1,m\pm1}^{i(e)} = 0$ [see Eq. (3)]. This stems inherently from the translational invariance of the external PSW field in the $xy$ plane. The second term $Q_s$ involving only $a_{l,m}^{i(j)}$ is contributed by the scattering from the other particles $j \neq i$, and the third term $Q_c$ is a cross term that involves the external incidence $a_{l,m}^{i(e)}$ and the scattering $a_{l,m}^{i(j)}$ simultaneously. It is straight to prove that $a_{l,m}^{i(j)}$ becomes negligibly small with respect to $a_{l,m}^{i(e)}$ as the inter-particle separation increases. Therefore, the cross term $Q_c$ is dominant for the dilute particle suspension.

Using the approximation $h_l(x) = \frac{(-i)^l e^{ix}}{ix} + \frac{(l+1)l}{2} \frac{(-i)^l e^{ix}}{x^2} + O\left(\frac{1}{x^2}\right)$ for $x \gg 1$, $a_{l,m}^{i(j)}$ in Eq. (6) can be reshaped as

$$a_{l,m}^{i(j)} = S_{l,m}^{ij(1)} \frac{e^{ikR_{ij}}}{kR_{ij}} + S_{l,m}^{ij(2)} \frac{e^{ikR_{ij}}}{\left(kR_{ij}\right)^2} + O\left(\frac{1}{\left(kR_{ij}\right)^2}\right), \quad (13)$$

where the coefficients

$$S_{l,m}^{ij(1)} = \sum_{l'} \left[ t_{l'}^j a_{l',0}^{j(e)} 4\pi i^{l-l'-1} \sum_{l''} C_{lml''(-m)}^{l'0} Y_{l'',-m}\left(\frac{\pi}{2}, \phi_{ij}\right) \right], \quad (14a)$$

$$S_{l,m}^{ij(2)} = \sum_{l'} \left[ t_{l'}^j a_{l',0}^{j(e)} 4\pi i^{l-l'} \sum_{l''} \frac{(l''+1)l''}{2} C_{lml''(-m)}^{l'0} Y_{l'',-m}\left(\frac{\pi}{2}, \phi_{ij}\right) \right]. \quad (14b)$$

with $\phi_{ij}$ denoting the azimuthal angle of $\hat{R}_{ij}$. Especially, one may prove that $S_{l,m}^{ij(1)} = 0$ if the particles are positioned on the node plane of the PSW field, since $a_{l',m'}^{j(e)} = 0$ for even $l'$ [see Eq. (3)] and $\sum_{l''} C_{lml''(-m)}^{l'0} Y_{l'',-m}\left(\frac{\pi}{2}, \phi\right) = 0$ for odd $l'$. Therefore, the total interaction force exerting on the sphere $i$ can be approximately written as

$$F_x^i + \mathrm{i}F_y^i \approx \sum_{j \neq i} \sum_l \frac{\mathrm{i}\rho_0}{4} \frac{1}{(kR_{ij})^n} \Big[\mu_{l+1,-1} a_{l,0}^{i(e)} \left(T_l^i e^{\mathrm{i}kR_{ij}} S_{l+1,-1}^{ij(n)} + T_l^{i*} e^{-\mathrm{i}kR_{ij}} S_{l+1,1}^{ij(n)*}\right)$$
$$+ \mu_{l+1,0} a_{l+1,0}^{i(e)} \left(T_l^i e^{-\mathrm{i}kR_{ij}} S_{l,1}^{ij(n)*} + T_l^{i*} e^{\mathrm{i}kR_{ij}} S_{l,-1}^{ij(n)}\right)\Big], (15)$$

where $n = 1$ and $n = 2$ correspond to the antinode and node planes, respectively. Note that here only the terms $S_{l,1}^{ij(n)}$ and $S_{l,-1}^{ij(n)}$ are involved, because $a_{l,m}^{i(e)}$ is nonzero only for $m = 0$ [see Eq. (3)].

To further simplify Eq. (15), we substitute the definition of the spherical harmonics $Y_{l,m}(\theta, \phi) = \zeta_l P_l^m(\cos\theta) e^{\mathrm{i}m\phi}$ into Eq. (14), with $\{P_l^m\}$ being the associated Legendre polynomials and $\zeta_l = \sqrt{[(2l+1)(l+1)!]/[4\pi(l-1)!]}$. This gives rise to

$$S_{l,1}^{ij(n)} = U_l^{j(n)} e^{-\mathrm{i}\phi_{ij}}, \quad S_{l,-1}^{ij(n)} = -U_l^{j(n)} e^{\mathrm{i}\phi_{ij}}. \tag{16}$$

with

$$U_l^{j(n)} = \sum_{l'} \left[ t_{l'}^j a_{l',0}^{j(e)} 4\pi \mathrm{i}^{l-l'-\delta_{n1}} \sum_{l''} \left[\frac{(l''+1)l''}{2}\right]^{\delta_{n2}} C_{l1l''(-1)}^{l'0} \zeta_{l''} P_{l''}^{-1}(0) \right]. \tag{17}$$

Combining Eq. (15) with Eq. (16), we derive a relatively simple and compact form

$$F_x^i + \mathrm{i}F_y^i = \sum_{j \neq i} \frac{A_{ij} \cos(kR_{ij} + \varphi_{ij})}{(kR_{ij})^n} e^{\mathrm{i}\phi_{ij}}. \tag{18}$$

The amplitude factor $A_{ij} = |Z^{ij(n)}|$ and the phase shift $\varphi_{ij} = \arg(Z^{ij(n)})$, where the complex quantity $Z^{ij(n)}$ is defined by

$$Z^{ij(n)} = -\mathrm{i}\frac{\rho_0}{2} \sum_l \left(\mu_{l+1,-1} a_{l,0}^{i(e)} U_{l+1}^{j(n)} T_l^i + \mu_{l+1,0} a_{l+1,0}^{i(e)} U_l^{j(n)} T_l^{i*}\right). \tag{19}$$

Interestingly, for Rayleigh particles $Z^{ij(n)}$ can be simplified further by considering the lowest two scattering channels only. This gives rise to $A_{ij} = \frac{\pi}{72} \Psi_0^2 \rho_0 (kD_i)^3 (kD_j)^3 f_{i,0} f_{j,0}$ and $\varphi_{ij} = \pi/2$ for the antinode plane, and

$A_{ij} = \frac{\pi}{32} \Psi_0^2 \rho_0 (kD_i)^3 (kD_j)^3 f_{i,1} f_{j,1}$ and $\varphi_{ij} = 0$ for the node plane, where $f_{i,0} = (\kappa_i - \kappa_0)/\kappa_i$ and $f_{i,1} = 2(\rho_i - \rho_0)/(2\rho_i + \rho_0)$ are the monopole and dipole scattering factors of the Rayleigh particle $i$, respectively. Here $\rho_i$, $\kappa_i$ and $D_i$ are the mass density, the bulk modulus and the diameter of the particle $i$, respectively, and $\kappa_0$ is the bulk modulus of the background fluid. The result responsible for the Rayleigh limit is consistent with that derived by Silva and Bruus [see Eqs. (22b) and (23b) in Ref. 24].

Below we give a summary on the property of the acoustic interaction among the sparsely distributed multiple particles. (**i**) The total force [see Eq. (18)] exerting on the sphere $i$ can be viewed as a vector addition of all pairwise interactions labeled by $ij$, since the quantity $Z^{ij(n)}$ in Eq. (19) depends on the scattering property of the particles $i$ and $j$ only. The additivity of the interaction does not hold naturally since the ARF is a second order quantity of the sound field. In this case, the conclusion stems from the negligible contribution of the interference among the scattering fields [see Eq. (12b)]. As a consequence, a multi-body system can be simply decomposed into a series of independent two-body problems, which will greatly simplify the computation of a dilute particle ensemble involving a large number of particles. (**ii**) For each pairwise interaction, the orientation of the force component depends on the azimuthal angle of $\hat{R}_{ij}$, i.e., $\phi_{ij}$. This means that the pairwise interaction is a 2D central force and thus is conservative, as long as the particle pair is tightly bounded on the node or antinode plane. Therefore, we can use a potential energy function to characterize the multi-particle system, considering the additivity of the pairwise force. (**iii**) As a whole, the pairwise interaction decays at a power exponential factor $(kR_{ij})^{-n}$, where $n = 1$ or $2$ corresponds to the case of the antinode plane or the node plane. Similar to the Coulomb force or the gravitational force, the acoustically mediated interaction decays moderately and is a long-range force. Besides, the cosine-like oscillation factor $\cos(kR_{ij} + \varphi_{ij})$ suggests a binding effect between the two particles: there exists a series of stable configurations occurring at $kR_{ij} + \varphi_{ij} =$ $(4N + 1)\pi/2$ with $N$ denoting an integer. The periodicity exhibited in the inter-particle separations of the bound states originate inherently from the distance dependent factor $a_{l,m}^{i(j)}$ [see Eq. (13)] involved in the dominant force component $Q_c$ [see Eq. (12c)]. Although a similar formula has been reported in optic systems [3,5,9], which involves the dipole-dipole interaction only, the present form of the pairwise acoustic interaction is not obvious since all scattering channels are considered here. This long-range force could be beneficial to designing and controlling thermodynamically stable 2D colloidal crystals.

### III. Numerical results and discussions

The sound-mediated interaction is closely related to the scattering property of the particles. In the following paper, the particles are assumed to be identical despite the fact that Eq. (18) can deal with an ensemble of particles with different geometry and material parameters. Specifically, here we consider polystyrene particles immersed in water, which carry rich scattering resonances at a wavelength comparable with the diameter $D$, as revealed in Fig. 2 by the magnitudes of the scattering matrix elements $|t_l|$. The material parameters involved are: the mass density $\rho = 1050 \text{kg/m}^3$, the longitudinal velocity $v_l = 2400 \text{m/s}$, and the transverse velocity $v_t = 1150 \text{m/s}$ for polystyrene; the mass density $\rho_0 = 1000 \text{kg/m}^3$ and the sound speed $c_0 = 1490 \text{m/s}$ for water. For the convenience of presentation, below all lengths are scaled by the operation wavelength $\lambda$, and the interaction force is scaled by $F_0 = E_0 S_0$, where $E_0 = \rho_0 k^2 \Psi_0^2 / 2$ is the energy density of a single plane wave, and $S_0 = \pi D^2 / 4$ is the cross-section area of the spherical particle. In all calculations, the cutoff of the angular quantum number $l_{max} = 8$ is used.

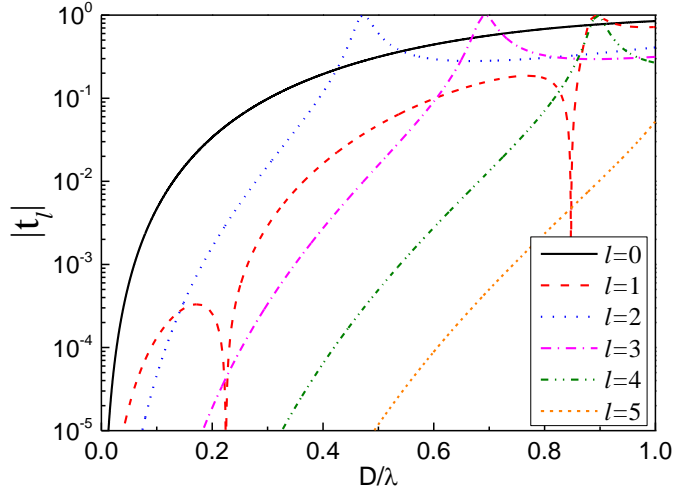

FIG. 2. (Color online) Magnitudes of the lowest six orders of the scattering matrix elements for a water-immersed polystyrene particle, plotted as a function of the dimensionless particle size $D/\lambda$.

## A. The system involving few particles

Consider the pairwise interaction first. According to Eq. (18), the normalized interaction force $F_N$ between a pair of particles $i$ and $j$ with a large separation $d$ can be expressed as

$$F_N(d/\lambda) = A_N(d/\lambda)^{-n}\cos(2\pi d/\lambda + \varphi). \tag{20}$$

where the amplitude factor $A_N = |Z^{ij(n)}|/F_0$, the phase shift $\varphi = \arg(Z^{ij(n)})$, and the power attenuation factor $n = 1$ or $2$ corresponds to the antinode or node plane. The positive (negative) sign of $F_N$ corresponds to a repulsive (attractive) force.

Figure 3 shows the pairwise interaction force plotted as a function of the dimensionless inter-particle separation $d/\lambda$, where the red dashed and black solid lines provide the comparative results obtained by the analytical formula Eq. (20) and the rigorous MST approach, respectively. Three particle sizes are considered, one close to the Rayleigh limit and the other two comparable with the acoustic wavelength. Depending on the stability of a single particle in the $z$ direction, the pair of particles are placed on the same node or antinode plane. It is observed that, the accuracy of Eq. (20) depends on the particle size and the inter-particle separation simultaneously. A remarkable deviation occurs at the first several wavelengths, especially for the big particles. As the inter-particle separation grows, the analytical results capture well the MST data for all particle sizes. Here we also give a brief comment on the energy barrier between two adjacent bound states. Consider the wave frequency 11MHz and the pressure amplitude 0.045MPa (experimentally used in Ref. 32). We find that the energy barrier, decaying as $1/d$ or $1/d^2$ again, is high enough to overcome the thermal effect. For instance, the energy barriers between two adjacent stable states (with $d/\lambda \sim 20$) are in the order of $10^2 k_B T$ for the case of $D/\lambda = 0.5$, and in the order of $10^5 k_B T$ for the case of $D/\lambda = 0.8$. Here $k_B$ is the Boltzmann constant and $T = 300K$ is the room temperature. This reveals the possibility of binding particles at large distance through acoustic waves.

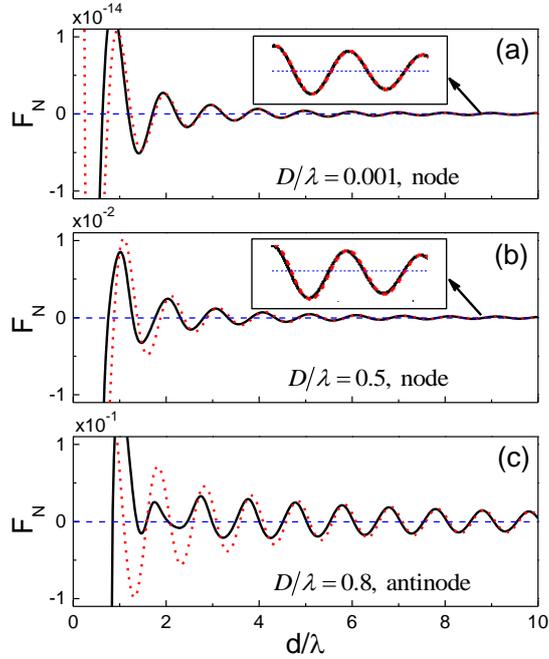

FIG. 3. (Color online) Distance dependence of the pairwise interaction force for three different particle sizes. The black solid and red dashed lines correspond to the results calculated by the rigorous MST method and the analytical formula Eq. (20), respectively. In (a) and (b), the insets indicate the enlarged forces for the large inter-particle distance.

As stated above, in addition to the well-defined power decay factor, the acoustic interaction can be described by two independent quantities, i.e., the dimensionless amplitude factor $A_N$ and the phase shift $\varphi$. Figure 4 shows $A_N$ and $\varphi$ for a pair of polystyrene particles located on the same node plane (the blue lines) or the antinode plane (the red lines) of the PSW field, plotted as a function of the particle size $D/\lambda$. It is observed that, for small particles ($D/\lambda \ll 1$), $A_N$ grows fast because of the rapidly enhanced scattering, accompanying with a nearly constant $\varphi$ as predicted in the Rayleigh limit. As $D/\lambda$ increases, $A_N$ becomes nonmonotonic and $\varphi$ varies sharply (where the latter suggests that the stable configuration is very sensitive to the particle size near the resonance). These features can be understood directly from the characteristics of the scattering matrix elements (see Fig. 2) and the incident coefficient $a_{lm}^{i(e)}$ [which is zero for either odd or even $l$, depending on the position of the particle, see Eq. (3)]. Besides, we have calculated the pairwise interaction by using the rigorous MST method, and fit the parameters $A_N$ and $\varphi$ according to Eq. (20). The fitted data agree well with the analytical ones again, as shown in Fig. 4 by the blue and red circles.

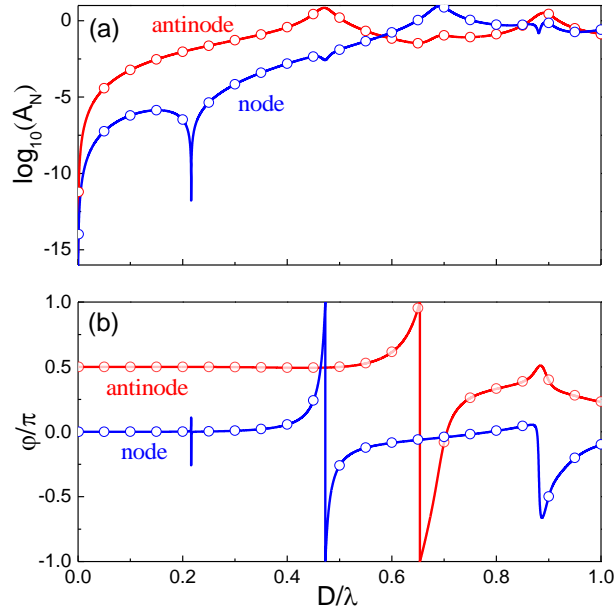

FIG. 4. (Color online) The amplitude factor $A_N$ (a) and the phase shift $\varphi$ (b) of the long-range acoustic interaction between a pair of polystyrene spheres located on the node plane (blue) or the antinode plane (red), plotted as the function of the dimensionless particle size $D/\lambda$. Here the lines are analytical results, and the circles are fitted from the rigorous MST method, respectively.

Now we check the vector additivity of the long-range acoustic interaction among multiple spherical particles. Without losing generality, we consider three identical polystyrene particles arranged into an equidistant linear array or a regular triangle array. As shown by the insets in Fig. 5, both configurations can be defined by the separation between adjacent particles. We consider the total force exerted on the third particle, which is along the $x$ direction according to the symmetry. It is easy to derived that for the linear and triangular configurations, the dimensionless total forces labeled by $F_{NL}$ and $F_{NT}$ can be expressed as

$$F_{NL}(d/\lambda) = F_N(d/\lambda) + F_N(2d/\lambda), \tag{21a}$$

$$F_{NT}(d/\lambda) = \sqrt{3} F_N(d/\lambda), \tag{21b}$$

where $F_N(d/\lambda)$ is the pairwise interaction in Eq. (20). In Fig. 5 we present the analytical results (red dashed lines) for both configurations, together with the data calculated by the rigorous MST method (black solid lines) for comparison. The particles with sizes $D/\lambda = 0.5$ and $D/\lambda = 0.8$ are considered and located on the node and antinode planes, respectively. Excellent agreements between the two approaches confirm well the vector additivity of the sound-mediated long-range interaction.

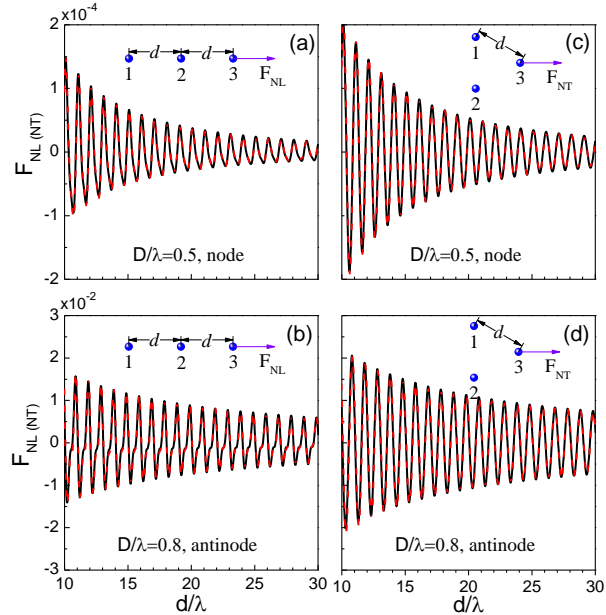

FIG. 5. (Color online) Verification of the vector additivity for the acoustically mediated long-range interaction, where (a) and (b) correspond to linear configurations, and (c) and (d) represent regular triangle configurations. The black solid and red

dashed lines correspond to the results calculated by the rigorous MST method and the analytical formula, respectively.

## B. The system involving a large number of particles

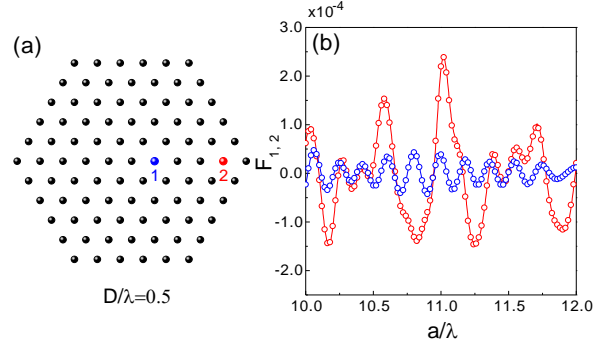

FIG. 6. (Color online) (a) A finite hexagonal lattice array of the polystyrene particles ($D/\lambda = 0.5$) located on the node plane of the PSW field. (b) The total interaction forces exerted on the particles 1 (blue) and 2 (red) labeled in (a), evaluated by the analytical formula (lines) and the rigorous MST method (circles).

Now we consider the systems involving many particles. Figure 6(a) shows 91 particles (diameter $D/\lambda = 0.5$) located on the node plane of the PSW field and arranged in a hexagonal lattice. We have calculated the total ARFs exerting on two particles (labeled by 1 and 2), one close to the center and the other near the boundary of the particle array, respectively. For these two particles, only the $x$-component is nonzero because of the symmetry. One hundred of configurations with different lattice constants ($a/\lambda$) are calculated by our analytical formula and the rigorous MST method, as shown in Fig. 6(b) by the solid lines and open circles. Excellent agreement between the two methods is obtained. In the MST approach, ~28 hours is paid to finish the calculation, in which solving the linear problem is time-consuming and requires large memory storage. In our analytic method, only ~1.0 second is cost, associated with a negligible memory requirement.

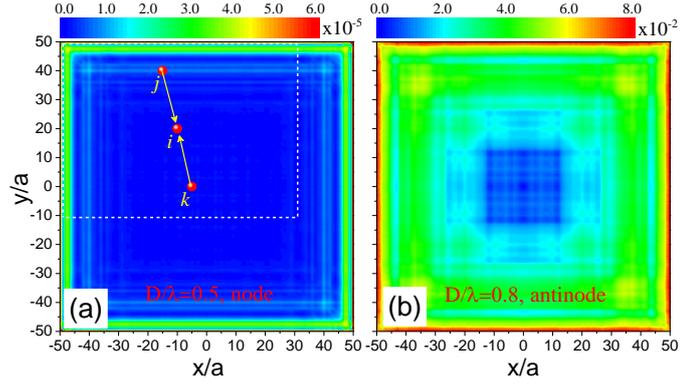

FIG. 7. (Color online) (a) The force pattern for 101x101 polystyrene particles (with the diameter $D/\lambda = 0.5$) arranged in a square lattice (with the lattice constant $a/\lambda = 10.36$), where the color indicates the amplitude of the total force exerting on the particle located at the corresponding position. The dashed rectangle frame indicates a particle region that contributes zero force to the particle $i$. (b) The same as (a), but for the particle size $D/\lambda = 0.8$ and the lattice constant $a/\lambda = 10.05$.

The significant optimization in computation time and memory cost also allows the analytic method to handle a system involving a huge number of particles, which is unattainable by the conventional MST approach. For instance, here we consider 101x101 particles arranged into a square lattice. The particle size is selected as $D/\lambda = 0.5$ or $D/\lambda = 0.8$. The lattice constant is taken as $a/\lambda = 10.36$ or $a/\lambda = 10.05$, one of the stable distances in the corresponding two-particle system. Obviously, the particle arrangement does not guarantee the stability of the whole system. For each particle, the total interaction exerted by the others has been calculated. In Figs. 7(a) and 7(b), we present the corresponding force amplitude distributions. Interestingly, Fig. 7(a) shows a much weaker force exerted on the inner particle than that force exerted on the particle near the boundary. Physically, the screening effect can be understood by the cancelation of many short-distance pairwise interactions. For instance, we consider the particle $i$ indicated in Fig. 7(a). It is easy to deduce that, the total force contribution from the particles inside the rectangle region (labeled by the dashed frame) is zero, e.g., the pairwise interaction contributed from the particle $j$ is canceled by that from the particle $k$, because of the inversion symmetry with respect to the particle $i$. The screening effect is also demonstrated in Fig. 7(b) for the case of $D/\lambda = 0.8$. Comparing with Fig. 7(a), the force decaying from the boundary is much slower, since the particles are located on the antinode plane now ($n = 1$).

## IV. Conclusion

Starting from a single prior scattering approximation, we have derived a compact analytical formula for the sound-mediated interaction among sparsely distributed particles. The formula, confirmed by the rigorous MST method, reveals that a multi-particle system can always be reduced to a two-particle problem, irrelevant to the size of particles. A combination of our analytical method with the MST approach could be helpful to further handle a big particle cluster with *arbitrary* inter-particle separations. The possibility of such new type of oscillatory long-range interactions in creating 2D colloidal crystals is also of great interest.

## Acknowledgements

This work is supported by the National Basic Research Program of China (Grant No. 2015CB755500); National Natural Science Foundation of China (Grant Nos. 11374233, 11534013, 11574233, and 11547310).

microparticles, J. Appl. Phys. **78**, 4845 (1995).